\documentclass{aa}  

\usepackage{graphicx}
\usepackage{txfonts}
\usepackage[colorlinks=true,linkcolor=black,citecolor=blue,urlcolor=blue,draft=false]{hyperref}
\begin{document}

   \title{An extended scattered light disk around AT Pyx
\thanks{Based on observations performed with VLT/SPHERE under program ID 099.C-0147(B) and  1100.C-0481(F), as well as VLT/NACO under program ID 1101.C-0092(D)}}

\subtitle{Possible planet formation in a cometary globule}

\author{C.~Ginski\inst{1,2}
\and R.~Gratton\inst{3}
\and A.~Bohn\inst{2}
\and C.~Dominik\inst{1}
\and S.~Jorquera\inst{5}
\and G.~Chauvin\inst{4,5}
\and J.~Milli\inst{4}
\and M.~Rodriguez\inst{6,1,5}
\and M.~Benisty\inst{4,5}
\and R.~Launhardt\inst{7}
\and A.~M\"uller\inst{7}
\and G.~Cugno\inst{8}
\and R.G.~van Holstein\inst{2}
\and A.~Boccaletti\inst{9}
\and G.\,A.\,Muro-Arena\inst{1}
\and S.~Desidera\inst{3}
\and M.~Keppler\inst{7}
\and A.~Zurlo\inst{10}
\and E.~Sissa\inst{3}
\and T.~Henning\inst{7}
\and M.~Janson\inst{7,11}
\and M.~Langlois\inst{12,13}
\and M.~Bonnefoy\inst{4}
\and F.~Cantalloube\inst{7}
\and V.~D'Orazi\inst{3}
\and M.~Feldt\inst{4}
\and J.~Hagelberg\inst{14}
\and D.~S\'egransan\inst{14}
\and A-M.~Lagrange\inst{4}
\and C.~Lazzoni\inst{3}
\and M.~Meyer\inst{15}
\and C.~Romero\inst{4}
\and T.O.B.~Schmidt\inst{9,16}
\and A.~Vigan\inst{13}
\and C.~Petit\inst{17}
\and R.~Roelfsema\inst{18}
\and J.~Pragt\inst{18}
\and L.~Weber\inst{14}
}

\institute{Anton Pannekoek Institute for Astronomy, University of Amsterdam, Science Park 904,1098XH Amsterdam, The Netherlands \email{c.ginski@uva.nl}
\and Leiden Observatory, Leiden University, 2300 RA Leiden, The Netherlands
\and INAF-Osservatorio Astronomico di Padova, Vicolo dell’Osservatorio 5, I-35122, Padova, Italy
\and Univ. Grenoble Alpes, CNRS, IPAG, 38000 Grenoble, France
\and Unidad Mixta Internacional Franco-Chilena de Astronom\'{i}a (CNRS, UMI 3386), Departamento de Astronom\'{i}a, Universidad de
Chile, Camino El Observatorio 1515, Las Condes, Santiago, Chile
\and Universidad de los Andes, M\'erida, Venezuela
\and Max Planck Institute for Astronomy, K\"{o}nigstuhl 17, 69117 Heidelberg, Germany
\and Institute for Particle Physics and Astrophysics, ETH Zurich, Wolfgang-Pauli-Strasse 27, 8093, Zurich, Switzerland
\and LESIA, Observatoire de Paris, Universit{\'e} PSL, CNRS, Sorbonne Universit{\'e}, Univ. Paris Diderot, Sorbonne Paris Cit{\'e}, 5 place Jules Janssen, 92195 Meudon, France
\and N\'ucleo de Astronom\'ia, Facultad de Ingenier\'ia y Ciencias, Universidad Diego Portales, Av. Ejercito 441, Santiago, Chile
\and Department of Astronomy, Stockholm University, SE-10691 Stockholm, Sweden
\and CRAL, CNRS, Universit\'{e} Lyon 1,Universit\'{e} de Lyon, ENS, 9 avenue Charles Andre, 69561 Saint Genis Laval, France
\and Aix Marseille Univ, CNRS, CNES, LAM, Marseille, France
\and Geneva Observatory, University of Geneva, Chemin des Maillettes 51, CH-1290 Sauverny, Switzerland
\and University of Michigan, Astronomy Department, 1085 S. University Ave., Ann Arbor, MI, 48109-1107, USA
\and Hamburger Sternwarte, Gojenbergsweg 112, 21029, Hamburg, Germany
\and DOTA, ONERA, Université Paris Saclay, F-91123, Palaiseau France
\and NOVA Optical Infrared Instrumentation Group, Oude Hoogeveensedijk 4, 7991 PD Dwingeloo, The Netherlands
}

   \date{}

 
  \abstract
   {}
   {To understand how the multitude of planetary systems that have been discovered come to be, we need to study systems at different evolutionary stages, with different central stars but also in different environments. The most challenging environment for planet formation may be the harsh UV radiation field of nearby massive stars which quickly erodes disks by external photo-evaporation. We have observed the AT\,Pyx system, located in the head of a cometary globule in the Gum Nebula, to search for signs of ongoing planet formation.}
   {We used the extreme adaptive optics imager VLT/SPHERE in Dual Beam Polarization Imaging Mode in H-band as well as in IRDIFS Extended mode (K12-band imaging and Y-H integral field spectroscopy) to observe AT\,Pyx in polarized light as well as total intensity. Additionally we employed VLT/NACO to observe the system in the L-band.}
   {We resolve the disk around AT\,Pyx for the first time in scattered light across multiple wavelengths in polarized light and total intensity. We find an extended ($\geq$126\,au) disk, with an intermediate inclination between 35$^\circ$ and 42$^\circ$. The disk shows a complex sub-structure and we identify 2 and possibly 3 spiral-like features. Depending on the precise geometry of the disk (which we can not unambiguously infer from our data) the disk may be eccentric with an eccentricity of $\sim$0.16 or partially self-shadowed.
   The spiral features and possible eccentricity are both consistent with signatures of an embedded gas giant planet with a mass of $\sim$1\,M$_{Jup}$. Our own observations can rule out brown dwarf companions embedded in the resolved disk, but are, however, not sensitive enough to confirm or rule out the presence of a gas giant.}
   {AT\,Pyx is the first disk in a cometray globule in the Gum Nebula which is spatially resolved. By comparison with disks in the Orion Nebula Cluster we note that the extension of the disk may be exceptional for this environment if the external UV radiation field is indeed comparable to other cometary globules in the region. The signposts of ongoing planet formation are intriguing and need to be followed up with either higher sensitivity or at different wavelengths.}


   \maketitle
%

\section{Introduction}

\begin{figure}
\center
\includegraphics[width=0.48\textwidth]{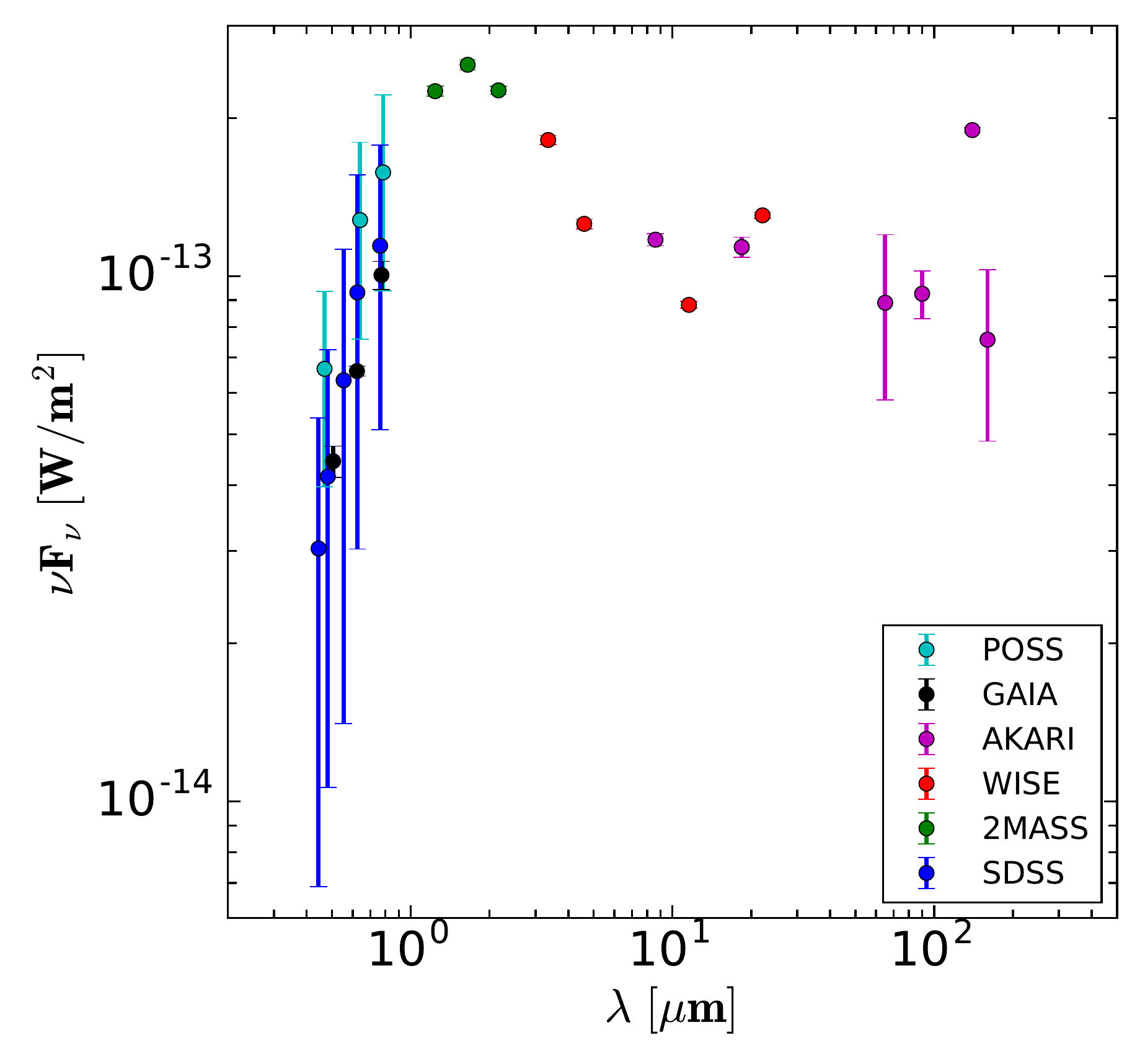} 
\caption{Observed spectral energy distribution of AT\,Pyx, based on photometric values (no de-reddening applied). The surveys from which measurements were taken are indicated by marker color. We note that the two longest wavelength AKARI data points may suffer from contamination, which may explain their discrepancy (\citealt{2010yCat.2298....0Y}). 
} 
\label{fig: sed}
\end{figure}

AT\,Pyx (=\,IRAS\,08267-3336, WRAY\,15-220) is a young intermediate mass star located in the Gum Nebula (\citealt{Herczeg2014}), a large HII region in the southern hemisphere.
Its distance was recently revised by Gaia to be 370$\pm$5\,pc (\citealt{2020arXiv201201533G,2020arXiv201203380L}).
Using the luminosity and temperature derived by \cite{Herczeg2014}, re-scaled to the new Gaia distance, in combination with \cite{2000A&A...358..593S} stellar isochrone models, we find a stellar mass of $1.5\,\pm\,0.1$\,M$_\odot$ and an age of 5.1$^{+1.5}_{-1.0}$\,Myr.\\
AT\,Pyx was first identified to be a pre-main sequence star by \cite{Pettersson1987}, who measured significant H$\alpha$ emission. We show the spectral energy distribution of the system in Figure~\ref{fig: sed}, assembled from various photometric catalogs. We note that \cite{Herczeg2014} measured an extinction of $A_V = 1.2\,mag$ toward the system. A clear infrared excess is present at wavelengths longer than 10$\mu$m. The dip in emission at the same wavelength is typical for a transition disk with an inner cavity. \cite{Garufi2018} find a fractional near infrared excess $F_{NIR}/F_* = 20.5\pm2.3\%$ and a far infrared excess of $F_{NIR}/F_* = 39.8\pm3.2\%$. The large far infrared excess is comparable to well known extended disk such as the ones around HD142527 (\citealt{Rodigas2014,Avenhaus2014}) or GG Tau (\citealt{McCabe2002,Itoh2014,Keppler2020}). The, compared to the far infrared excess, lower near infrared excess is typical for disks bright in scattered light (\citealt{Garufi2018}).\\
In the All-Sky Automated Survey for SuperNovae (ASAS-SN) catalog (\citealt{2019MNRAS.486.1907J}), a strong V-band variability of AT\,Pyx was found with an amplitude of 1.45\,mag and time scales on the order of days. This may indicate that AT\,Pyx is a so-called "dipper" star in which inner disk material is obscuring the star periodically (see e.g. \citealt{2015AJ....149..130S, 2015A&A...577A..11M}).\\
The system is located in the head of a cometary globule (see Figure~\ref{fig:sphere_images}, left), i.e. a dense region of molecular gas with a fading tail (\citealt{Hawarden1976}). The origin of the cometary globules are still not entirely clear. 
They may have been caused by the shock-wave of the high mass companion of $\zeta$\,Pup which went supernova roughly 1.5\,Myr ago (\citealt{Sahu1988}), or they may be caused by the interaction with the intense radiation field of the Vela OB association (\citealt{2009MNRAS.393..959C}).
Cometary globules have been found to be sites of enhanced star formation, likely triggered by the same event that created the Gum Nebula (\citealt{Bhatt1993}).\\
Due to the UV environment caused by the nearby Vela OB association, it is thought that circumstellar disks should have a shortened lifetime compared to low mass star forming regions. 
Indeed \cite{Kim2005} found that out of 11 PMS stars located in cometary globules that they observed, only one showed a significant infrared excess. 
This lead them to the conclusion that the typical disk lifetime should be shorter than 5\,Myr due to external photo-evaporation.
However, the cometary globules observed by \cite{Kim2005} are located closer to the O4 star $\zeta$\,Pup, than the cometary globule that hosts AT\,Pyx, thus the UV background radiation should be lower in the case of AT\,Pyx.\\
\cite{Guedel2010} detected ionized Neon in the mid infrared towards AT\,Pyx, indicative of stellar x-ray or EUV radiation hitting the surface of a circumstellar disk. This may be an indication of an enhanced EUV external radiation field, but might also be explained by the chromospheric activity of AT\,Pyx itself. \\
In this study we present high resolution scattered light observations of the AT\,Pyx system in the near infrared with VLT/SPHERE (\citealt{Beuzit2019}) and VLT/NACO (\citealt{Lenzen2003,Rousset2003}). 


\section{Observations and data reduction}

In the following we describe the observation setup and the data reduction of all observing epochs.

\subsection{SPHERE observations}

\subsubsection{Polarimetric observations}

AT\,Pyx was observed on May 15, 2017 with SPHERE/IRDIS (\citealt{2008SPIE.7014E..3LD}) in dual polarization imaging mode (DPI, \citealt{Langlois2014,deBoer2020,vanHolstein2020}). Observations were carried out in H-band with a coronagraph blocking the central region of the system (92.5\,mas inner working angle, \citealt{2011ExA....30...39C}) and the instrument de-rotator operating in field-stabilized mode. The observations were part of the SPHERE guaranteed time survey of nearby T\,Tauri stars. We give the detailed observation setup and conditions in table~\ref{tab: obs_summary}.\\
The data reduction was performed with the IRDIS pipeline for Accurate Polarimetry (IRDAP, \citealt{vanHolstein2020}) using default parameters. The details of the data reduction are described in \cite{vanHolstein2020}. We show the final Q$_\phi$ image in Figure~\ref{fig:sphere_images}, bottom-right panel.
The Q$_\phi$ image contains all the azimuthally polarized signal as positive values and all radial polarized signal as negative values, see \cite{deBoer2020} and \cite{2019ApJ...872..122M} for a detailed explanation. For a single star illuminating a nearly face-on circumstellar disk we expect in principle only azimuthally polarized light and thus the Q$_\phi$ image should give the majority of the scattered light signal received from the system. 
We additionally show the initial combined and flux calibrated Stokes Q and U images as well as the complementary U$_\phi$ image in appendix~\ref{appendix:pol}.

\subsubsection{Intensity observations}

We observed AT~Pyx in SPHERE IRDIFS mode on April 18, 2018 within the SPHERE Consortium Guaranteed Time Observations. The IRDIFS-EXT mode uses simultaneously the Integral Field Spectrograph (IFS: \citealt{Claudi2008}) over the wavelength range 0.95-1.65~$\mu$m at a spectral resolution of $R\sim 30$ (field of view 1.77 arcsec square), and the dual band imager IRDIS \citep{Dohlen2008} for the K1 and K2 narrow bands at 2.09 and 2.22~$\mu$m, respectively, over a wider field of view of $\sim 10$ arcsec square. We used the ALC\_Ks coronograph \citep{Boccaletti2008} with a field mask having a radius of 120 mas. Observing conditions were excellent with a median DIMM seeing FWHM of 0.82 arcsec and a coherence time of 9.4 ms that provided an average Strehl ratio of 0.62 in the K-band on this optically faint (Gmag = 13.45, \citealt{Gaia}) target. The observations were carried out in pupil stabilized mode and the total field rotation angle was 77.6 degrees. The observing sequence included in addition to the science exposure (a total of 4608 sec on target), observations with the star offset by about 0.5 arcsec with respect the coronagraph for flux and point spread function calibration, observations with a bi-dimensional sinusoidal pattern imprinted on the deformable mirror to provide faint replicas of the stellar image for fine centering (\citealt{Beuzit2019}), and background sky observations.\\
Data were reduced using the SPHERE DRH pipeline \citep{Pavlov2008} and additional routines available at the SPHERE Data Center \citep{Delorme2017}. The data were then analysed using the SPECAL routines \citep{Galicher2018}, that includes simple rotation and sum, angular differential imaging (ADI: \citealt{Marois2006}), TLOCI \citep{Marois2014}; and Principal Component Analysis (PCA: \citealt{Soummer2012, Amara2012}). For IFS we also used additional routines based on the PCA method simultaneously in space and wavelength coordinates developed at INAF-Osservatorio Astronomico di Padova \citep{Mesa2015}.

\subsection{NACO observations}
The NACO observations were obtained on January 19, 2019 as part of the ISPY (Imaging Survey for Planets around Young stars, \citealt{2020A&A...635A.162L}) observation campaign. observations were carried out in the L' filter with the L27 objective and a pixel scale of 27.19\,mas.. Since AT\,Pyx is relatively faint in the L-band, no coronagraph was used to block the central star. To accurately sample the variable sky background in the L-band the individual frame exposure time was set to 0.2\,s. The total integration was 46\,min.  
The observation setup and conditions are summarized in table~\ref{tab: obs_summary}.\\
The data were reduced using the IPAG-ADI pipeline \citep{2012A&A...542A..41C}. Data calibration (flat-fielding, bad pixels and sky removal) was performed as a first step on all available cubes. To reduce computation time, sub-frames of 90 $\times$ 90 pixels (FoV of $ 2.4"\times2.4"$) were extracted. Finally, these frames were re-centered and bad frames (poorly saturated, overly extended PSF) were removed to obtain a final master cube used for the data reduction.\\
The pipeline allows for the use of multiple flavours of ADI algorithms, namely classical ADI (cADI), smart ADI (sADI), radial ADI (rADI), all described in \cite{2012A&A...542A..41C}, as well as Locally Optimized Combination of Images \citep[LOCI, ][]{2007ApJ...660..770L} and Principal Component Analysis \citep[PCA, ][]{Soummer2012}. The use of multiple ADI techniques, together with the LOCI and PCA reduction, allows for comparison and consistency on the results obtained. For sADI and LOCI, we followed a similar configuration as \cite{2012A&A...542A..41C}, with a $FWHM=4.5$ pixels and a separation criteria of $0.75 \times FWHM$ at the companion separation. For the PCA method \citep[e.g.][]{Soummer2012}, observations were reduced using three different number of modes ($k=1,5,20$) and masking the image outside of a radius of 60 pixels.

\begin{figure*}
\center
\includegraphics[width=0.98\textwidth]{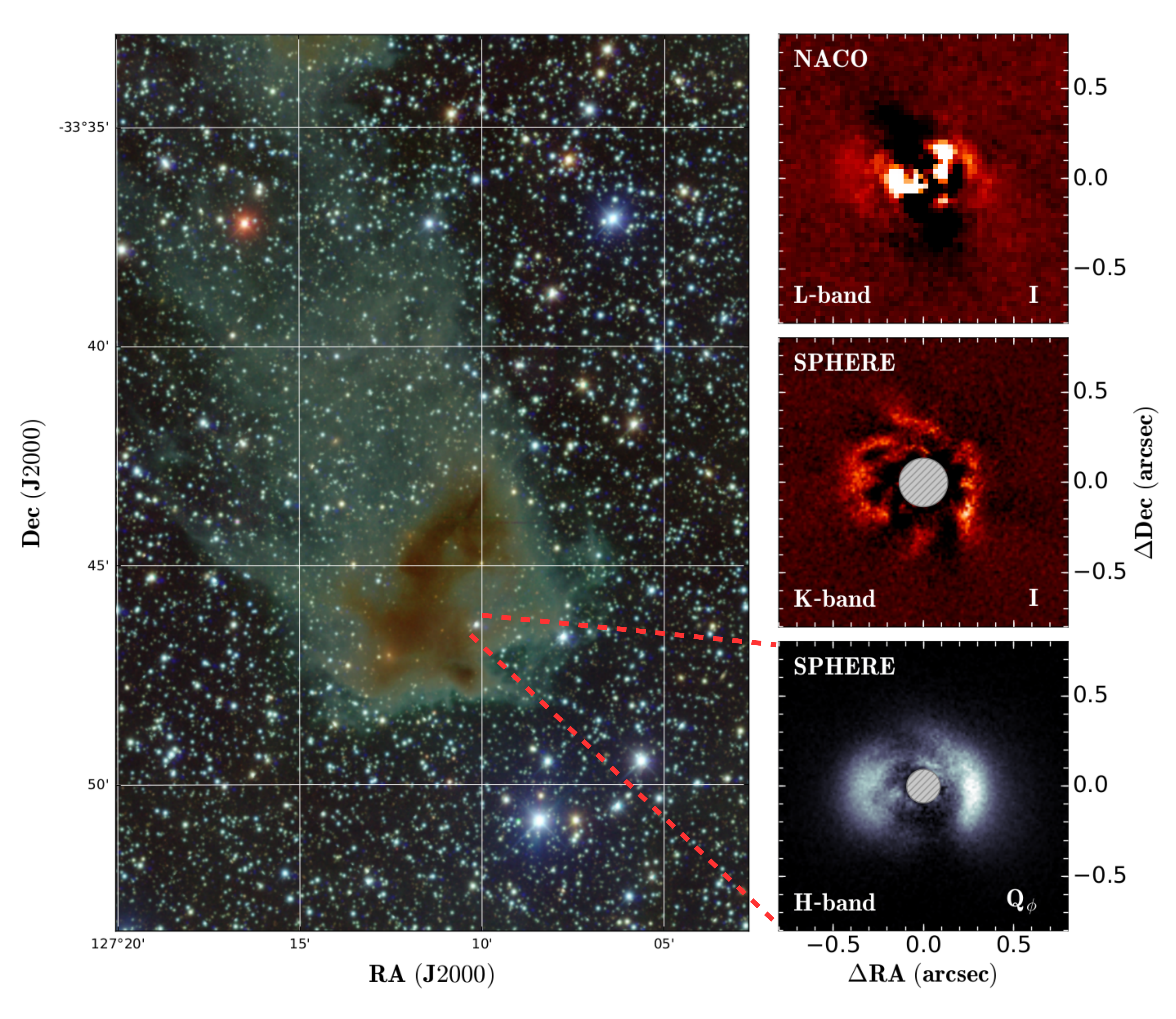} 
\caption{\emph{Left:} Dark Energy Cam Plane Survey (DECaPS, \citealt{Schlafly2018}) optical multi-band image of the head of the cometary globule CG-22 in the Gum Nebula. The red dashed lines indicate the position of AT\,Pyx. \emph{Right:} From top to bottom: NACO L-band ADI image of the disk around AT\,Pyx; SPHERE K-band ADI image and SPHERE H-band polarized light image.  
} 
\label{fig:sphere_images}
\end{figure*}

\begin{table*}
 \centering
 \caption{Observation setup and observing conditions.}
  \begin{tabular}{@{}lcccccccc@{}}
  \hline 
 Date  		& Instrument & Mode & Filter	    & Coronagraph	   & DIT [s] & \# Frames & Seeing [arcsec] & $\tau_0$ [ms]\\
 \hline 
 15-05-2017	& SPHERE & DPI &    BB\_H		& YJH\_ALC	& 96      & 12		&	0.75	&	- \\
 18-04-2018	& SPHERE & IRDIFExt &    K1/K2+IFS		& Ks\_ALC	& 96	  & 45		&	0.82	&	9.4	\\
 19-01-2019	& NACO   & int &    L'		    & none	    & 0.2	  & 13800	&	0.80	&	10.3	\\
 
 \hline

\hline\end{tabular}
\label{tab: obs_summary}
\end{table*}


\section{The scattered light disk}

\begin{figure*}
\center
\includegraphics[width=0.9999\textwidth]{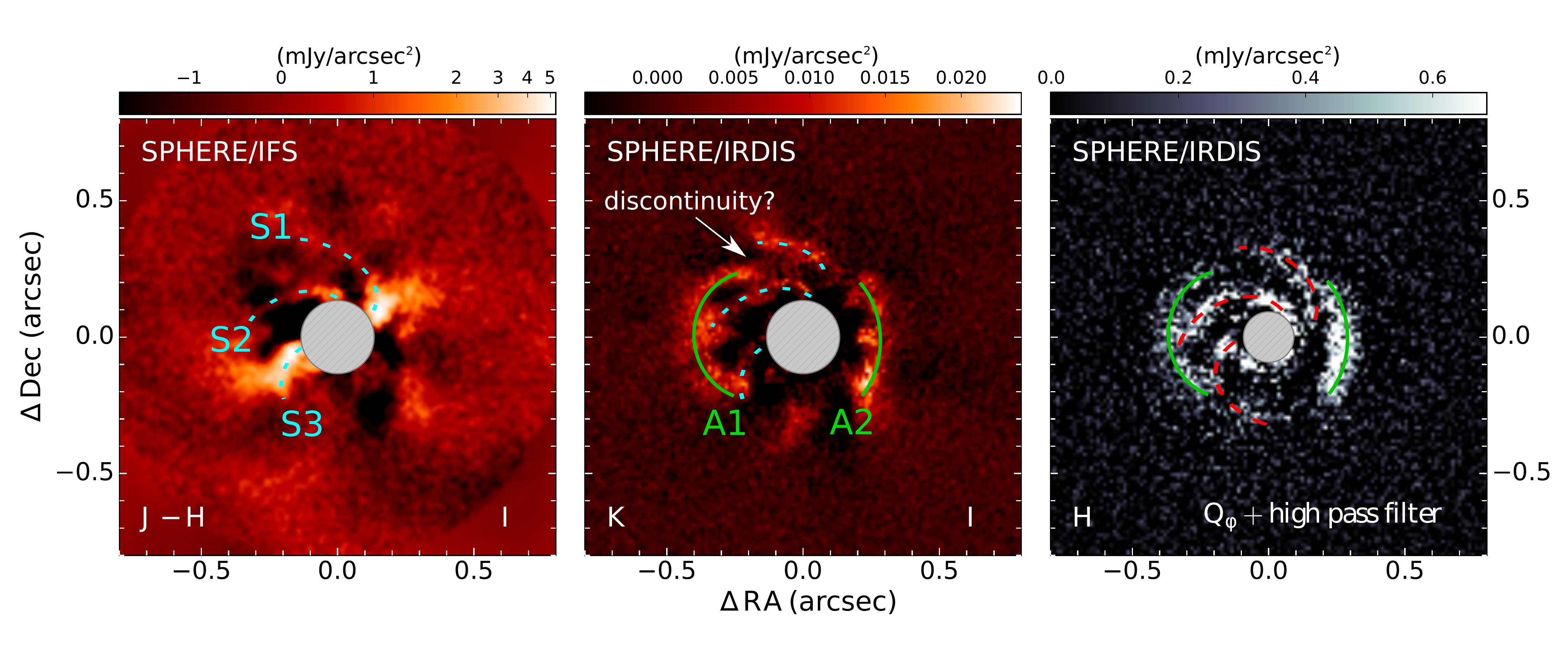} 
\caption{(\emph{Left:}) SPHERE/IFS image after angular differential imaging was applied to remove the stellar light component.
(\emph{Center:}) SPHERE/IRDIS K-band image after angular differential imaging.
(\emph{Right:}) SPHERE/IRDIS H-band image after polarization differential imaging. A high pass filter was applied to highlight high spatial frequency features and make the image more comparable to the K-band and the IFS images.
We mark similar features in all three images with dashed lines for the "spiral-like" features S1 - S3 and with continuous lines for the "arc-like" features A1 and A2. In the center panel we specifically point out the discontinuity between feature S1 and A1 in the North.
We note that the flux calibrations are not absolute values due to the varying throughput of the different data processing methods (this also explains the negative flux values). They can be used to follow variations within each panel but are not comparable between panels.
} 
\label{fig: features}
\end{figure*}

\begin{figure}
\center
\includegraphics[width=0.48\textwidth]{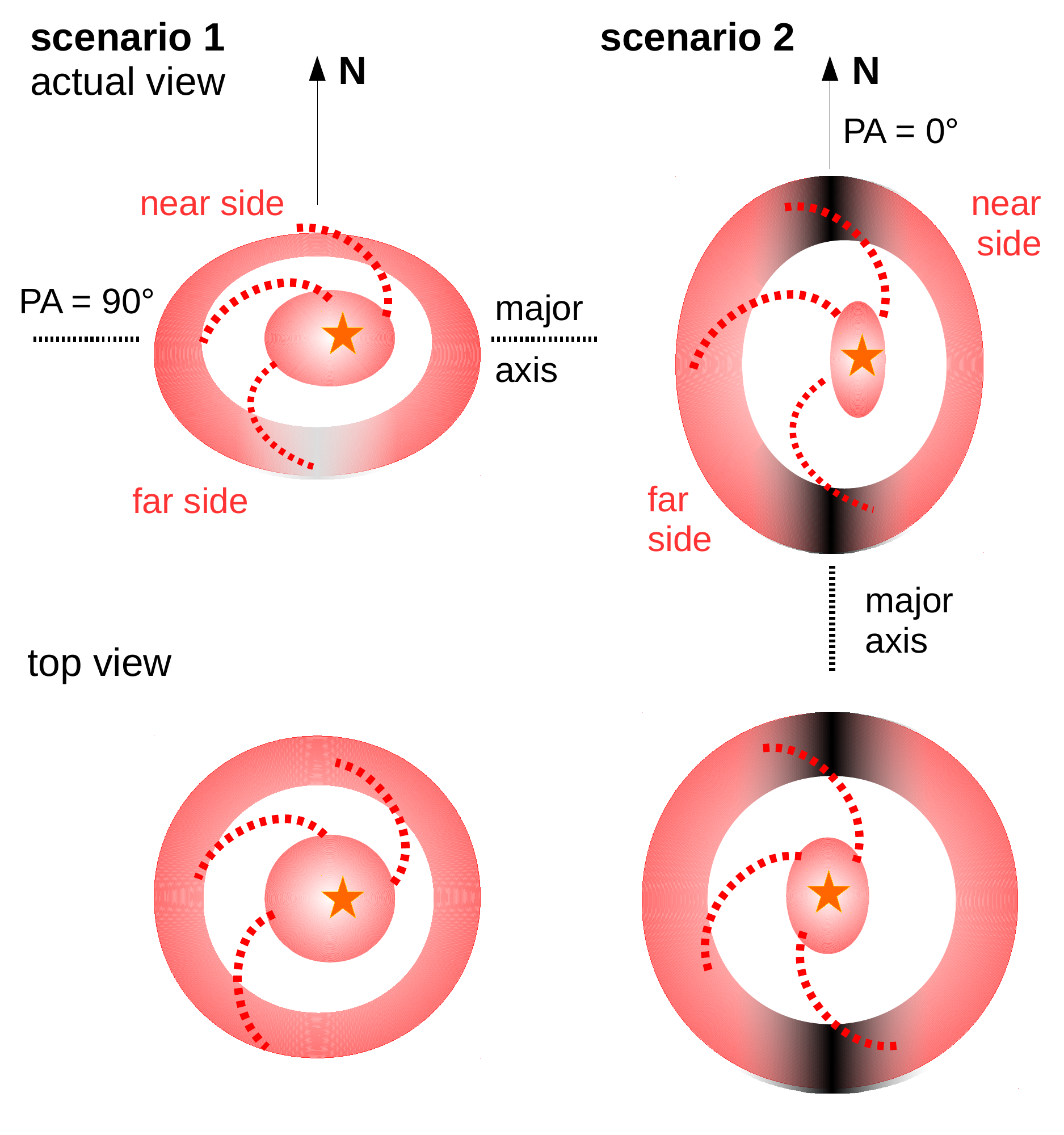} 
\caption{Schematic of the two scenarios to explain the scattered light features of the disk around AT\,Pyx. Scenario 1 (left) proposes an eccentric disk with a position angle of the major axis of 90$^\circ$. Scenario 2 (right) proposes a circular disk with a position angle of 0$^\circ$ and a misaligned inner disk, which shadows the ansae of the visible outer disk. For both scenarios we show a schematic of the actual view (top row) and of the de-projected top-down view (bottom row).
} 
\label{fig: disk-cartoon}
\end{figure}

Our multi-wavelength scattered light observations reveal an extended disk with complex sub-structures. In Figure~\ref{fig: features} we highlight the most prominent structures in the SPHERE IFS and IRDIS data sets. We find two arc-like structures to the East (A1) and West (A2) of the central star, best visible in the IRDIS data sets (and to a lesser degree in the NACO data set in Figure~\ref{fig:sphere_images})\footnote{We investigate why the ring-like features do not appear in the IFS data in appendix~\ref{appendix:simulation}.}. These features may present an outer ring of the disk. We furthermore find three spiral-like features starting in the North and East of the innermost disk area that we resolve and winding counter-clockwise (features S1-S3). These potential spirals are detected in all SPHERE images, however their position and shape changes slightly between the three data sets presented in Figure~\ref{fig: features}. This is likely an effect of the post processing method applied to the IFS and IRDIS K-band data. ADI processing is well known to alter the shape of extended features due to self-subtraction (see e.g. \citealt{Milli2012,Ginski2016}). The H-band polarized light image does not suffer from this effect and thus traces the morphology of the features most reliably. To allow for an easier cross-identification of the various features between the data sets, we show the IRDIS K-band data as well as the NACO L-band data, with the SPHERE H-band polarimetric data overlayed in contours in figures~\ref{fig:sphere_sphere_overlay} and \ref{fig:naco_sphere_overlay} in appendix~\ref{appendix:ident}\\
The interpretation of the observed structures depends strongly on the projection effects due to the disk inclination. Given our data we propose two scenarios for the disk inclination and position angle which we also illustrate in figure~\ref{fig: disk-cartoon}:\\
\\
\emph{Disk position angle of $\sim$90$^\circ$}\\
\\
The features A1 and A2 could represent the ansae of an inclined disk. They appear particularly bright in polarized light, which could be an effect of the polarized phase function peaking at scattering angles of $\sim$90$^\circ$. If this is the case, then we would expect the near side of the disk to be in the North, since we receive in all data sets more flux from the North compared to the South, consistent with smaller scattering angles. If this interpretation is correct then we should be able to fit features A1 and A2 with a single ellipse (assuming that both features trace the same height structure in scattered light). As is visible in Figure~\ref{fig: features} the curvature of feature A1 appears to be larger than of feature A2, which is inconsistent with this picture. However it may be that we misidentified feature S1 as an individual spiral-like feature and that instead it is a continuation of feature A2. In this case the discontinuity between the Eastern tip of S1 and the Northern tip of A1 that we point out in the center panel of Figure~\ref{fig: features} is puzzling.
Such a discontinuity might be explained if a shadow is cast on the outer ring from an inner (unresolved) disk with a relative inclination, similar to the case of the HD\,142527 system (see e.g. \citealt{Marino2015}). The corresponding shadow in the South might then not be detected due to the general low signal in that region. It might also be possible that feature S1 is a spiral arm that is significantly lifted above the disk surface profile. In this case the discontinuity might simply be a projection effect. A combination of both explanations for the appearance of the Northern side of the disk is also possible, i.e., a spiral arm with a significantly increase scale height compared to the surrounding disk, which casts shadows on the surrounding structures.\\
If we fit features A1 and A2 (ignoring feature S1) with an ellipse, then we find an inclination of the disk of $\sim$42$^\circ$ and a semi-major axis of the outer features of $\sim$126\,au. However to allow for a simultaneous fit of both features with the same ellipse the center needs to be offset from the central star by 55\,mas. Such an offset along the major axis can not be explained by projection effects and would indicate that the disk is eccentric with an eccentricity of $\sim$0.16. \\
\\
\emph{Disk position angle of $\sim$0$^\circ$}\\
\\
In order to allow for a non-eccentric outer ring, traced by features A1 and A2, we consider that the disk position angle might rather be close to $\sim$0$^\circ$. In this scenario it is possible to fit both features with an ellipse with an offset along the minor axis of the disk. This offset can be explained by the projection of a flared and inclined disk with the near side in the West (see e.g. \citealt{deBoer2016}). Feature A2 appears brighter than A1 in the SPHERE H-band and the K-band images\footnote{We note that in principle only the SPHERE H-band image preservers the (polarized light) photometry, while this is not the case for the ADI processed K and L-band images which suffer from possibly complex self-subtraction effects.}. This may then be explained by the scattering phase function, if A2 is the near side and thus is seen under smaller angles. This is consistent with feature A2 being closer to the stellar position in the 2d-plane than feature A1. Assuming this position angle we find a similar inclination of the disk to the previous scenario of $\sim$35$^\circ$ with a semi-major axis of the outer features of $\sim$145\,au. However, we caution that in this scenario the ansae of the disk are not seen, so the semi-major axis and the inclination are less constrained.\\
If we assume this orientation, then the lack of flux in the South and South-West in the polarized light data is puzzling, since these areas would be seen under close to 90$^\circ$ scattering angles and the degree of polarization should be near maximum compared to any other area in the disk. Thus if this is the correct disk orientation we require that the South and South-West but also the North and North-East are shadowed by interior structures. In the North this may be due to the spiral-like features S1 and S2. While S3 seemingly extends into the South in the left panel of Figure~\ref{fig: features}, it does not extend visibly to the South-West, where we would expect the peak of the polarized scattered light signal. However, the polarized light H-band data show signal close to the coronagraph in the North and South, best visible in Figure~\ref{fig: features}, right panel. If this signal traces an inner disk that is misaligned with respect to the outer disk, traced by features A1 and A2, then this may produce the required shadowing along the ansae.\\
\\
Based on our data sets, we are unable to unambiguously determine which of the proposed scenarios represents the correct interpretation of the disk morphology. Longer wavelength data, e.g. mm continuum or line emission observations are needed to disentangle the large and small scale morphology of the disk. We do however note, that the strong optical variability of AT\,Pyx might favor a rather inclined inner disk, consistent with the shadowing scenario and a disk position angle of 0$^\circ$.

\section{Planet detection limits}

\begin{figure}
\center
\includegraphics[width=0.48\textwidth]{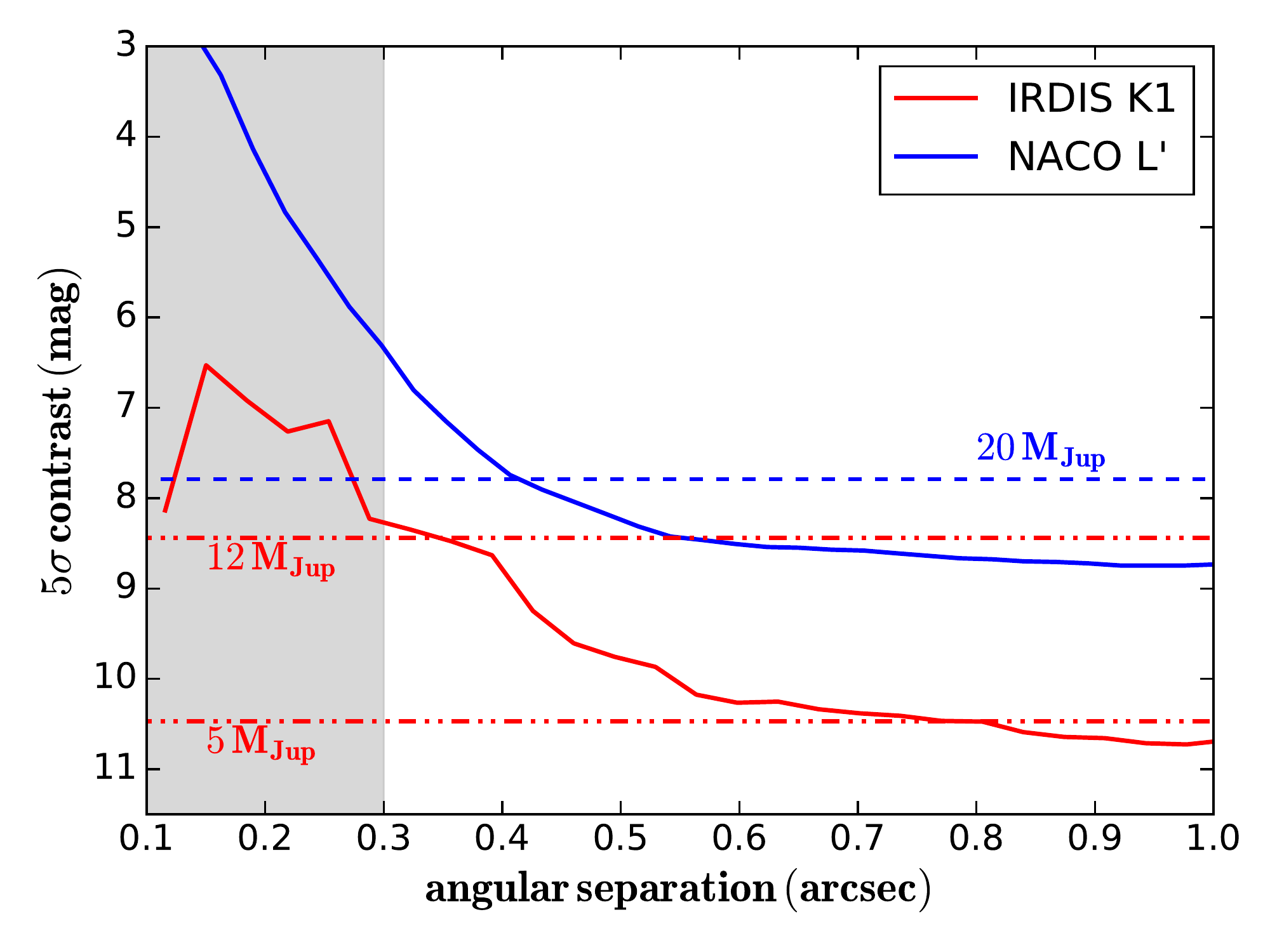} 
\caption{Achieved contrast in the SPHERE K-band (red) and the NACO L-band (blue) observations. We indicate the contrast for 5\,M$_{Jup}$ and 12\,M$_{Jup}$ objects in K-band and a 20\,M$_{Jup}$ object in L-band. We used AMES-DUSTY models to convert the contrast to mass limits assuming a system age of 5\,Myr. The grey shaded area marks the inner cavity of the scattered light disk inside of features A1 and A2.
} 
\label{fig: detection-limits}
\end{figure}

The spiral features discussed in the previous section might be an indication of a perturber embedded in the disk, e.g. a forming planet or a low-mass stellar component. Both such objects would be expected to be brightest relative to the central star at long wavelengths. Additionally the optical depth of the disk decreases towards longer wavelengths allowing to observe more embedded objects.\\
In neither the SPHERE K-band data nor the NACO L-band data we find unambiguous signatures of a point source in or directly outside the disk. We note that due to the complex sub-structure in combination with the ADI processing it is possible to produce false-positive or false-negative results (see e.g. the case of LkCa\,15, \citealt{Currie2019}). In Figure~\ref{fig:sphere_images} there is a tantalizing point-like feature present in the NACO data at a separation of roughly 0.1\arcsec{} to the South-West of the star. However, this signal has no counterpart in the higher contrast SPHERE K-band or the SPHERE IFS data. We thus assume it is an ADI-distorted disk structure or a remnant of the stellar PSF.\\
Using the SPHERE and NACO data we computed contrast limits, shown in Figure~\ref{fig: detection-limits}. To transform these contrast limits into mass detection limits we used AMES-DUSTY models (\citealt{Allard2012}), with a system age of 5\,Myr. To compute absolute magnitudes from the contrast limits we utilized the 2MASS K-band magnitude of the system of 9.044$\pm$0.021\,mag for the SPHERE data. For the NACO L-band data at a wavelength of 3.8\,$\mu m$ we interpolate between  
the WISE W1 magnitude at 3.4\,$\mu m$ and the W2 magnitude at 4.6\,$\mu m$. We calculate a value of 7.75$\pm$0.03\,mag.\\
We find that outside of 1\arcsec{} (i.e. outside of disk features A1 and A2), we are sensitive to 5\,M$_{Jup}$ planets in the SPHERE data and 20\,M$_{Jup}$ brown dwarfs in the NACO data. 
Inside of the disk cavity (i.e. inside of features A1 and A2, grey shaded area in Figure~\ref{fig: detection-limits}), we achieve significantly lower contrast, due to the coinciding disk structures. We can firmly rule out brown dwarf or low-mass stellar companions, but are not sensitive to massive gas giant planets.\\
We note that presented detection limits do not take into account extinction effects. Given that the system is still young, it is conceivable that planets inside the disk cavity are (partially) enshrouded by circumplanetary material. This may to a lesser degree also be the case for planets outside of the disk. Extinction may lead to shallower detection limits than shown in Figure~\ref{fig:sphere_images}. For a recent, detailed discussion of these effects we refer to \cite{2021arXiv210305377A}.

\section{Discussion and conclusions}

Our observations spatially resolve for the first time a disk around a T\,Tauri star in the Gum Nebula. We find a very extended disk out to at least 126\,au with complex sub-structure. AT\,Pyx appears to be an exceptional system in more than one respect. It is not only one of the few young stellar systems embedded in a cometary globule that still hosts a circumstellar disk, but also the radial extent of the disk is larger than expected for a disk which may be undergoing external photo-evaporation. \cite{Eisner2018} studied the mm emission of disks in the Orion Nebula Cluster (ONC), i.e., a high UV environment, and found that the majority of their sample showed outer radii smaller than 20\,au. Masses in these disks are also typically small, with average values of $\sim$ 3\,M$_{Jup}$ in the close vicinity (0.03\,pc) of the Trapezium cluster (\citealt{Eisner2008,Mann2010,Mann2014}). From our scattered light observations we can not draw direct conclusions about the mass of the disk surrounding AT Pyx. We note however that similar extended disks in scattered light have masses up to 50 times higher than the value reported for the photo-evaporated ONC disks (\citealt{Garufi2018}).
The presence of such an extended and potentially massive disk might be explained by the young age of AT\,Pyx. \cite{Clarke2007}
found that disks should be significantly eroded by external photo-evaporation after 1-2\,Myr. The age estimate of AT\,Pyx is only slightly older ($\sim$5\,Myr). Given typical model uncertainties for the ages of young T\,Tauri stars (e.g. \citealt{Pecaut2012}), it might be possible that AT\,Pyx is still younger and only now in the process of loosing its outer disk.
However, we note that we only have indirect evidence that AT\,Pyx is indeed located in a high background radiation field, the strongest of which is its location in the head of a cometary globule. \\
The sub-structure that we resolve shows two or possibly three spiral features\footnote{Following the discussion in the previous section we consider feature S1 ambiguous}. Spiral density waves in the gas are a common prediction for planets embedded in disks.
Massive planets, with masses approaching the disk thermal mass or exceeding it, typically excite non-linear density waves and can thus drive multiple spiral arms (\citealt{Goodman2001, Rafikov2002, Juhasz2015, Dong2015, Bae2016}). In particular \cite{Zhu2015} and \cite{Fung2015} find configurations with three spiral arms that are similar in appearance to the observations of AT\,Pyx. This typically requires planets at least more massive than Neptune (\citealt{Dong2017}). \\
In scenario 1 that we outlined for the disk geometry, with a position angle of $\sim$90$^\circ$, we find that the disk must be eccentric. Such an eccentricity may also well be explained by an embedded massive planet. \cite{Zhang2018} ran an extensive grid of hydrodynamic models and found that for their higher mass planets ($>$1\,M$_{Jup}$) and small disk aspect ratios at the planet position of h/r = 0.05 to 0.07, the planet opens a significantly eccentric gap in the gas (e $>$ 0.15). Since we trace micron sized particles that are well coupled to the gas we can expect to trace such an eccentricity also in scattered light (Muro-Arena et al., in prep.). To estimate if such a high planet mass may be consistent with our observations we used the relation between planet mass, spiral scattered light peak contrast and aspect ratio found by \cite{Dong2017}. Since feature S2 is the most unambiguous one, we measured its peak brightness near the launching point in the H-band polarized light data and compared it to the azimuthal average of the disk flux at the same separation. We found a spiral-to-disk contrast ratio of 2.29. Assuming aspect ratios of 0.05 and 0.07 and using equation 14 from \cite{Dong2017} this yields planet masses of 0.7\,M$_{Jup}$ and 1.1\,M$_{Jup}$, respectively. This is well consistent with the required planet mass to open an eccentric gap. Due to the large distance of the AT\,Pyx system and the additional confusing disk signal, the thermal radiation of such a planet is unfortunately below our detection threshold.\\
Alternatively, if scenario 2 holds true with a disk position angle of 0$^\circ$, then a strong misalignment of inner and outer disk might also indicate the presence of a perturbing companion, which may be of planetary or stellar nature, similar as in the case of the disk around HD\,142527 (\citealt{2016A&A...590A..90L,2018MNRAS.481.3169P}). A relative to the line of sight strongly inclined inner disk may also explain the strong optical variability of AT\,Pyx, especially if its highly structured, such that dust clumps can occult the star. Given the location in the head of CG-22, it may also be possible that a relative disk misalignment between inner and outer disk is caused by material infall, similar as was recently inferred in the case of the SU\,Aur system (\citealt{2021ApJ...908L..25G}). In this case it may be possible that the dynamic structures that we observe in the outer disk were also triggered by an infall event, without the need of a perturbing compact object. \\
The morphology of the AT\,Pyx system is intriguing and may well present the first detection of dynamic signs of ongoing planet formation in an externally photo-evaporated disk. To confirm this inference, ALMA observations in the sub-mm are needed to determine the basic disk geometry and mass. High resolution CO line observations may enable us to trace the kinematic signatures of the spiral features seen in scattered light. While a direct detection of the thermal radiation of the embedded planets is challenging with current ground or space based instrumentation, it may well be possible to observe accretion signatures which can be more easily disentangled from disk scattered light (e.g. \citealt{2019NatAs...3..749H}).

\begin{acknowledgements}
    We would like to express our appreciation for the extraordinary contribution of the late France Allard to the field of exoplanet atmospheres. The model isochrones that we use in this study are a direct result of her work. 
    This work has been supported by the project PRIN-INAF
2016 The Cradle of Life - GENESIS- SKA (General Conditions in Early Planetary
Systems for the rise of life with SKA
and PRIN-INAF 2019 "Planetary systems at young ages (PLATEA)"
This work has made use of data from the European Space Agency (ESA) mission
{\it Gaia} (\url{https://www.cosmos.esa.int/gaia}), processed by the {\it Gaia}
Data Processing and Analysis Consortium (DPAC,
\url{https://www.cosmos.esa.int/web/gaia/dpac/consortium}). Funding for the DPAC
has been provided by national institutions, in particular the institutions
participating in the {\it Gaia} Multilateral Agreement.
SPHERE was designed and built by a consortium made of IPAG (Grenoble, France), MPIA (Heidelberg, Germany), LAM (Marseille, France), LESIA (Paris, France), Laboratoire Lagrange (Nice, France), INAF–Osservatorio di Padova (Italy), Observatoire de Genève (Switzerland), ETH Zurich (Switzerland), NOVA (Netherlands), ONERA (France) and ASTRON (Netherlands) in collaboration with ESO.  SPHERE was funded by ESO, with additional contributions from CNRS (France), MPIA (Germany), INAF (Italy), FINES (Switzerland) and NOVA (Netherlands).  
Additional funding from EC's 6th and 7th Framework Programmes as part of OPTICON was received (grant number RII3-Ct-2004-001566 for FP6 (2004–2008); 226604 for FP7 (2009–2012); 312430 for FP7 (2013–2016)). We acknowledge the Programme National de Planétologie (PNP) and the Programme National de Physique Stellaire (PNPS) of CNRS-INSU, France, the French Labex OSUG@2020 (Investissements d’avenir – ANR10 LABX56) and LIO (Lyon Institute of Origins, ANR-10-LABX-0066 within the programme Investissements d'Avenir, ANR-11-IDEX-0007), and the Agence Nationale de la Recherche (ANR-14-CE33-0018) for support.
\end{acknowledgements}

%
%

\bibliographystyle{aa}
\bibliography{myBib} 

\begin{appendix}
\section{SPHERE polarimetric images}
\label{appendix:pol}

\begin{figure*}[!h]
\center
\includegraphics[width=0.98\textwidth]{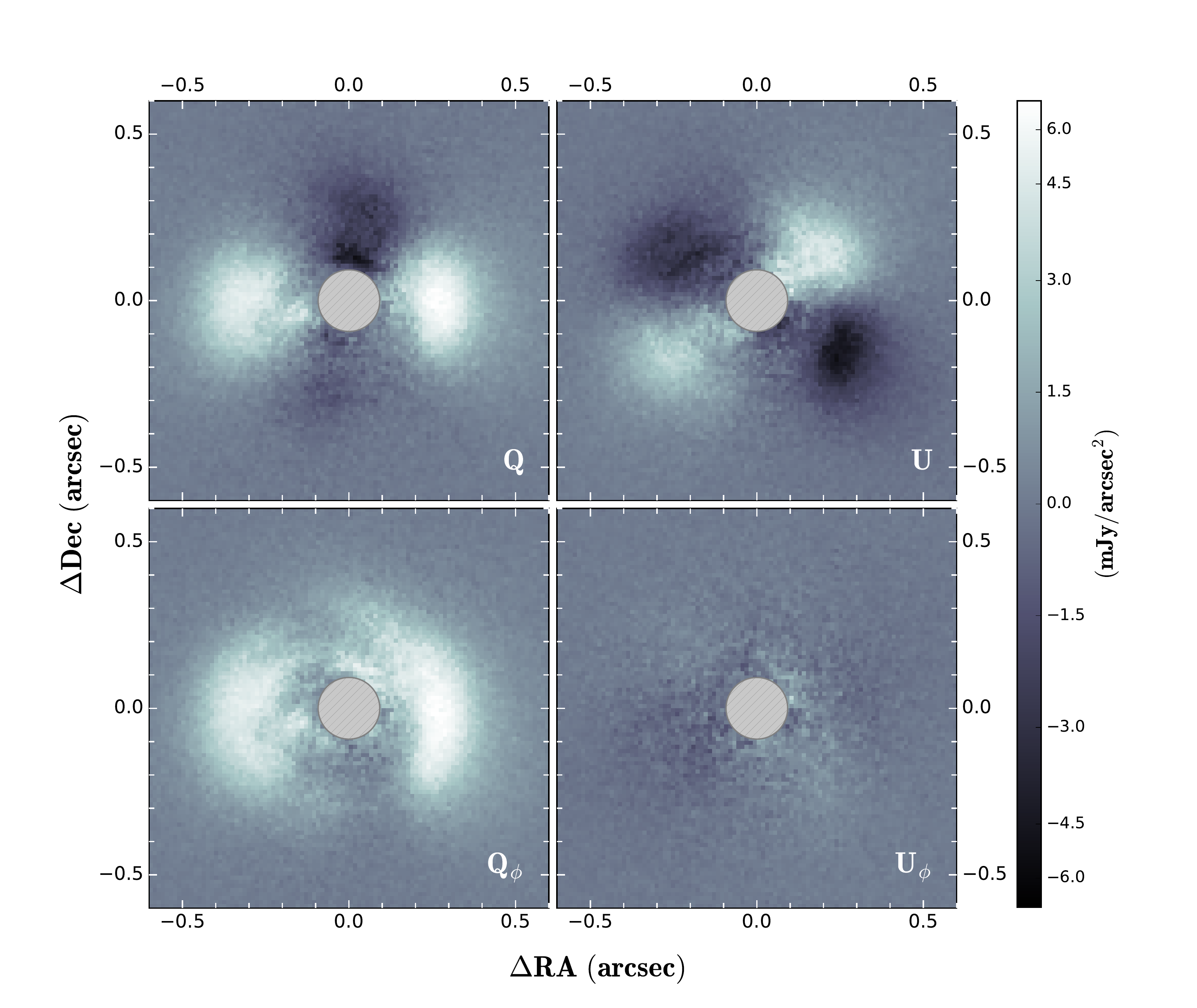} 
\caption{SPHERE/IRDIS H-band polarimetric images of AT\,Pyx. To highlight all features of the disk the color map is partially logarithmic, starting at 4 $mJy/arcsec^2$. All images are shown on the same color scale. Images are aligned with true north up and east to the left. The grey, hashed area marks the coronagraphic mask that was blocking the stellar light.  
} 
\label{fig:sphere_pol}
\end{figure*}

In Figure~\ref{fig:sphere_pol} we present the Stokes Q and U, as well as the derived Q$_\phi$ and U$_\phi$ images. The flux calibration was carried out by measuring the flux of the central star in the non-coronagraphic flux calibration images, taken at the beginning and end of the observation sequence. To convert pixel counts to physical units we used the 2MASS H-band magnitude of AT\,Pyx as reference. The U$_\phi$ image is showing very low flux levels compared to the Q$_\phi$ image, which indicates that the polarization signal is heavily dominated by single scattering of the light of the central star. This is expected for a circumstellar disk seen under a relatively low inclination, compatible with our analysis.

\section{Cross identification of sub-structures}
\label{appendix:ident}
In order to better cross-identify sub-structures between the different observational data sets, we present contour overlays of the polarized light H-band data on top of the SPHERE K-band and NACO L-band total intensity data. The structures A1 and A2, as well as S1, S2 and S3 match closely in position and shape between the SPHERE data sets. The NACO data set has lower spatial resolution and contrast close to the star and does not detect features S1 or S3. It does detect the extended ring-like features A1 and A2. We note that feature A1 in the NACO data has a "forked" appearance on the northern end, which is similar to what is seen in the PDI contours and is there caused by the merging of feature S2 with feature A1. In that sense feature S2 might be regarded as marginally detected in the NACO data as well. 

\begin{figure}[!h]
\center
\includegraphics[width=0.48\textwidth]{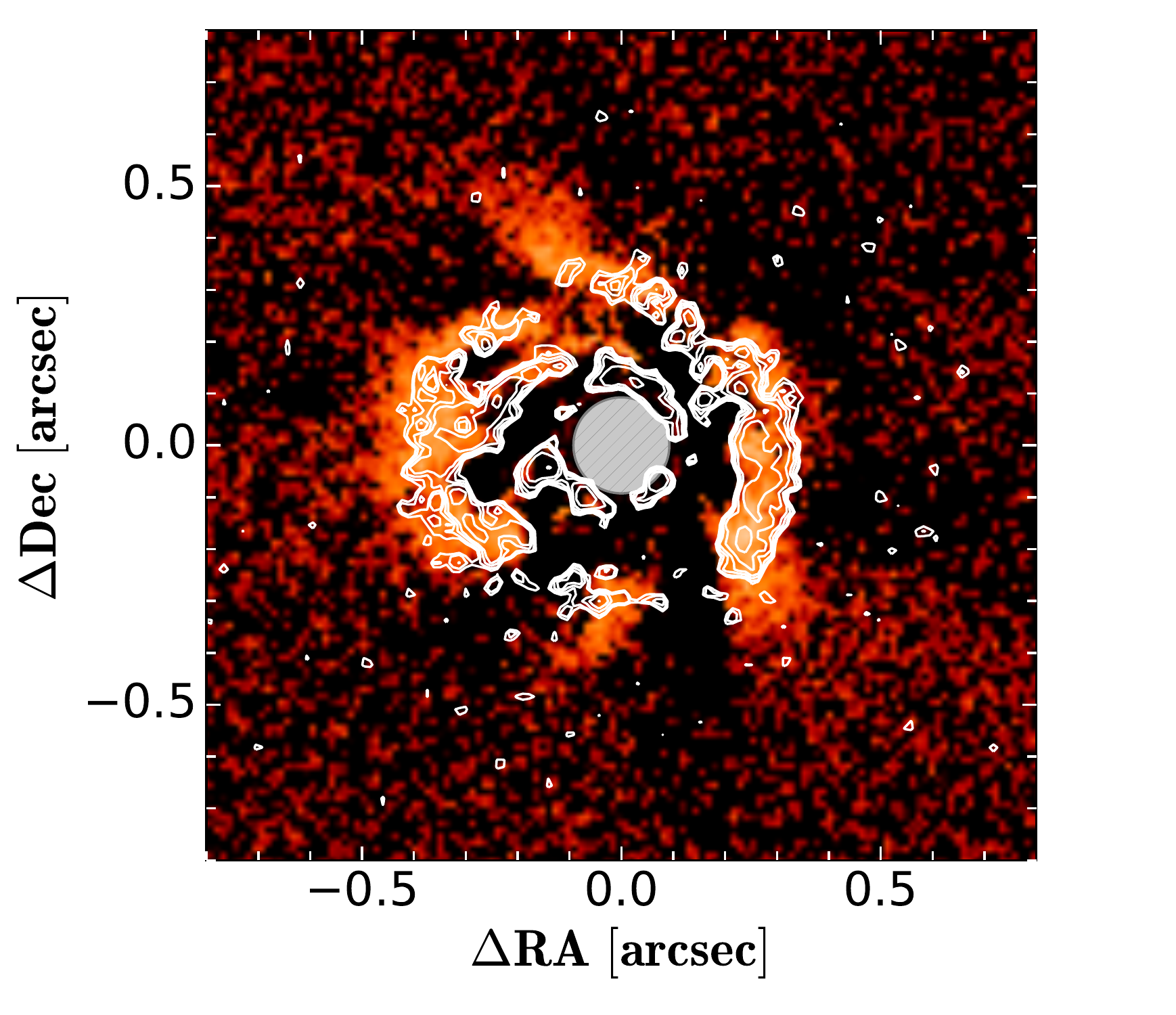} 
\caption{SPHERE/IRDIS K-band image after ADI post-processing (same as in figure~\ref{fig: features}), shown in a logarithmic color map. The white overlayed contours are drawn from the SPHERE/IRDIS H-band Q$_\phi$ image after high-pass filtering (same data set as shown in figure~\ref{fig: features}). The contours associated with the features S1, S2 and S3, match closely in shape and position to the position of the same features as seen in total intensity. We note that we display the size of the H-band coronagraph with the grey, hashed circle. 
} 
\label{fig:sphere_sphere_overlay}
\end{figure}

\begin{figure}[!h]
\center
\includegraphics[width=0.53\textwidth]{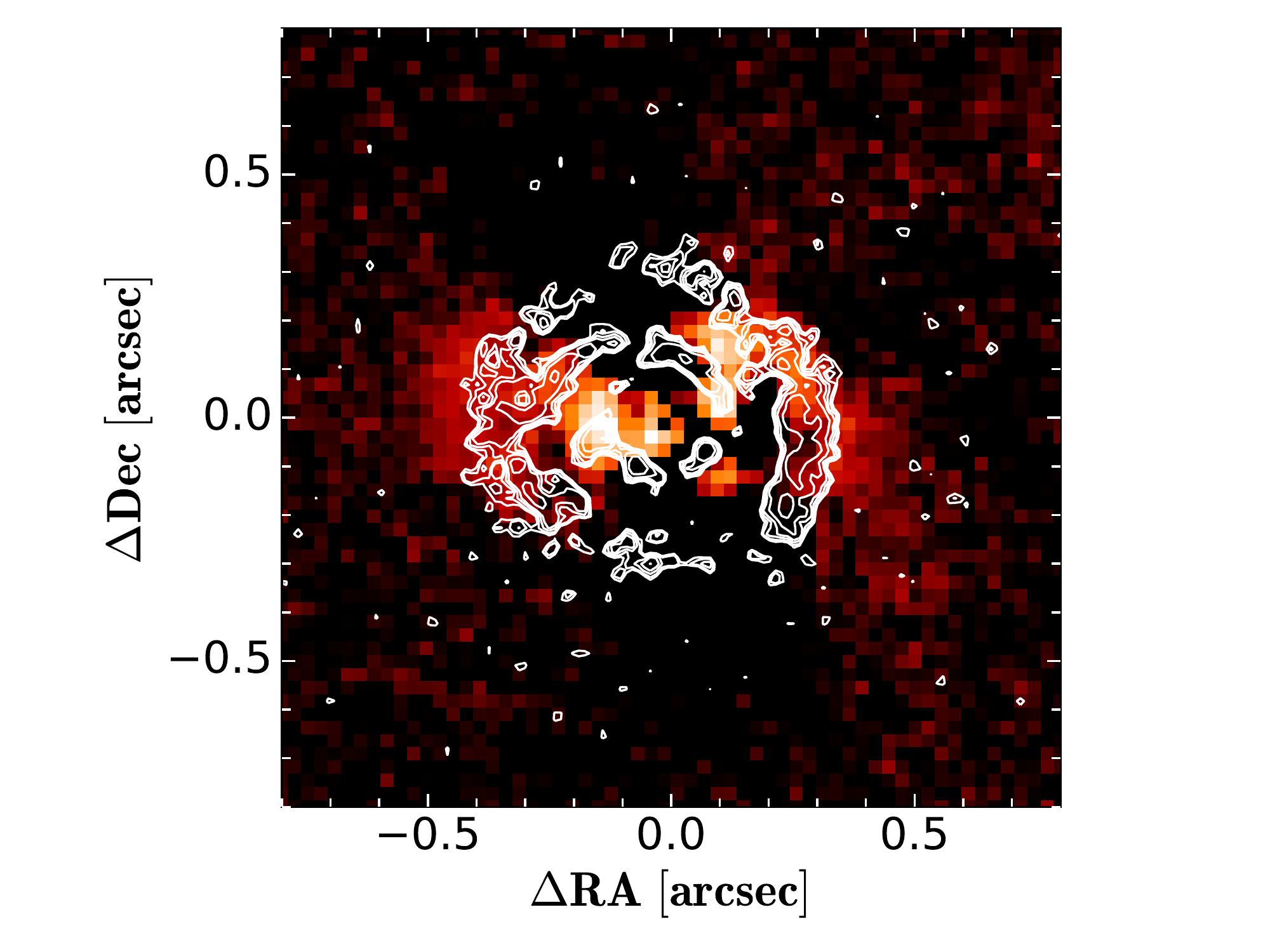} 
\caption{NACO L-band image after ADI post-processing (same as in figure~\ref{fig:sphere_images}), shown in a logarithmic color map. The white overlayed contours are drawn from the SPHERE/IRDIS H-band Q$_\phi$ image after high-pass filtering (same data set as shown in figure~\ref{fig: features}). The contours associated with the features A1 and A2, match closely in shape and position to the position of the same features as seen in total intensity. 
} 
\label{fig:naco_sphere_overlay}
\end{figure}

\section{Simulated ADI on PDI data}
\label{appendix:simulation}
\begin{figure}[!h]
\center
\includegraphics[width=0.53\textwidth]{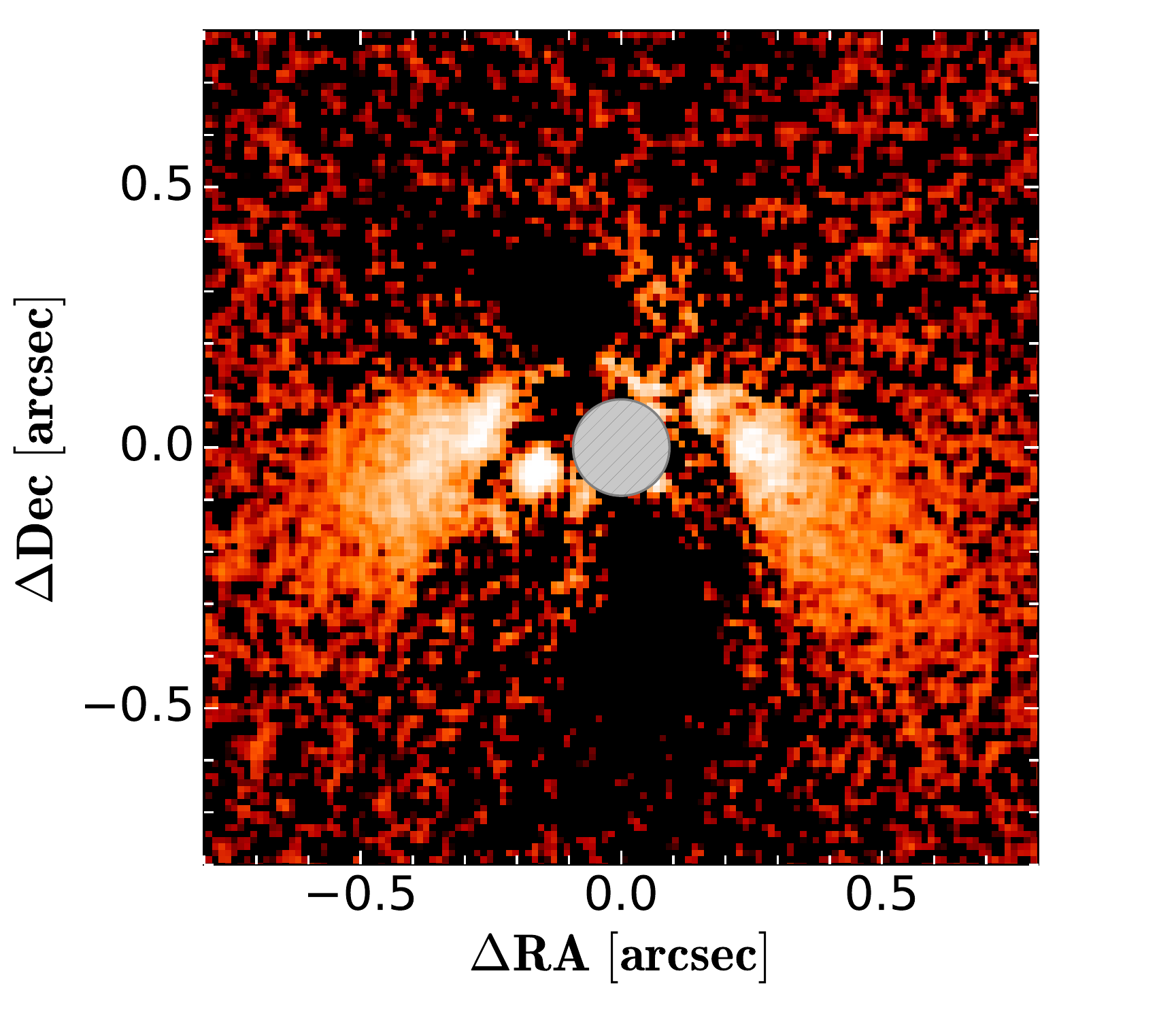} 
\caption{IRDIS H-band polarized light data set after post-processing by simulated classical ADI. For the field rotation during the ADI simulation the parallactic angles of the IRDIS K-band data set were used. The ring-like features A1 and A2 are strongly suppressed.
} 
\label{fig:simulated ADI}
\end{figure}

To test whether the non detection of the features A1 and A2 in the IFS data might be a post-processing effect, we simulated classical ADI post processing on the IRDIS PDI H-band data. The result is shown in figure~\ref{fig:simulated ADI}.
The ring-like features A1 and A2 are strongly suppressed since they extended predominantly in the azimuthal direction. The spiral-like features S1, S2 and S3 are to some degree visible in the data, but at low SNR. We note that a perfect match with the IFS data set is not to be expected since the IFS data shows total intensity while we only used the polarized light as input for the simulation. Additionally the noise characteristic of IRDIS and the IFS is fundamentally different since the IFS has a higher thermal background due to a lack of cooling. Nevertheless it is clear that ADI post-processing strongly suppresses the ring-like features in the data.

\end{appendix}

\end{document}